\documentclass[11pt]{article}

\usepackage[margin=1in]{geometry}
\usepackage[numbers,sort&compress]{natbib}
\usepackage{amsmath,amssymb}
\usepackage{graphicx}
\usepackage{booktabs}
\usepackage{array}
\usepackage{microtype}
\usepackage[colorlinks=true,citecolor=blue,linkcolor=blue,urlcolor=blue]{hyperref}

\title{Proportion Outcomes in Interrupted Time Series Analysis: Why Both Boundedness and Autocorrelation Matter}
\author{Ariel Linden\\[4pt]
\small University of California, San Francisco\\
\small Department of Medicine, Division of Clinical Informatics \& Digital Transformation\\
\small \href{mailto:ariel.linden@ucsf.edu}{ariel.linden@ucsf.edu}}
\date{}

\begin{document}

\maketitle

\begin{abstract}
\noindent\textbf{Background:} Proportions are common outcomes in interrupted time series analysis (ITSA), but their bounded support and the potential serial dependence among observations present distinct modeling considerations. Common analytical approaches may address one of these features without addressing the other. This teaching article provides practical guidance for recognizing these
issues and selecting and evaluating models for proportion outcomes in
ITSA.

\noindent\textbf{Methods:} Two published healthcare ITSA studies were reanalyzed to illustrate these complementary problems. The first contained weekly proportions concentrated near the boundaries of the unit interval and had originally been analyzed using segmented linear regression. The second contained monthly proportions lying within the unit interval and had been analyzed without reported assessment or adjustment for autocorrelation. The studies were reanalyzed using models appropriate for their
respective outcome distributions and, where indicated, temporal
dependence.

\noindent\textbf{Results:} In the first example, segmented linear regression produced six fitted values below zero. Overall, 35 of 66 confidence intervals extended beyond the permissible range of the outcome. Fractional logistic regression constrained the fitted conditional mean to the unit interval while accommodating observed values of zero and one. In the second example, beta autoregressive regression identified significant first-order serial dependence ($\rho_1=0.263$, 95\% CI 0.093--0.432; $P=.002$). After modeling this dependence, conditional residuals showed no evidence of remaining autocorrelation (Ljung--Box $Q_{12}=15.59$; $P=.211$).

\noindent\textbf{Conclusions:} Boundedness and serial dependence are separate modeling considerations in ITSA of proportion outcomes. Methods that account for autocorrelation do not necessarily respect the permissible range of a proportion; conversely, models appropriate for bounded outcomes do not necessarily account for serial dependence. Investigators should therefore assess both features explicitly and select an analytical approach appropriate to both the outcome and its temporal dependence.
\end{abstract}


\section{Introduction}

Interrupted time series analysis (ITSA) is widely used in clinical, health
services, and public health research to evaluate interventions when
randomization is infeasible \cite{campbell1966,shadish2002,linden2015}.
ITSA has been argued to have strong internal validity, primarily through
its control over regression to the mean
\cite{campbell1966,shadish2002,linden2007}, and generally has good
external validity, particularly when the unit of measurement is at the
population level or when results can be generalized to other units,
treatments, or settings \cite{shadish2002,linden2004}. In the
single-group design, observations collected at regular intervals before
and after an intervention are used to estimate the pre-intervention
trend and changes in the level and trend of the outcome following the
intervention \cite{linden2015}. Valid inference depends not only on
specifying these intervention-associated changes but also on using a
model appropriate for the outcome and its temporal dependence
\cite{lopezbernal2018,turner2020}.

Many outcomes analyzed using ITSA are proportions or percentages boun\-ded
by zero and one. Examples include the proportion of patients receiving
a recommended treatment, prescriptions meeting specified criteria, or
procedures complying with a clinical guideline. Nevertheless, proportion
outcomes are sometimes analyzed using methods developed for continuous
unbounded outcomes. Linear regression, for example, does not constrain
the conditional mean to the unit interval and can produce fitted values
or confidence limits below zero or above one, particularly when
observations lie near either boundary. Conventional autoregressive integrated moving-average (ARIMA) models can
address serial dependence but likewise do not inherently constrain
fitted values to the permissible range. Thus, accounting for the
time-series structure does not by itself address the bounded nature of
a proportion outcome.

Serial dependence must also be considered separately. Failure to account
adequately for autocorrelation can result in incorrectly estimated
standard errors, confidence intervals, and hypothesis tests
\cite{linden2026a,linden2026b,turner2021,bottomley2023}. Turner et al.\
found that autocorrelation was acknowledged in only 63\% of published
ITSA series and that an estimate of the autocorrelation coefficient was
reported in only 1\% \cite{turner2020}. The data underlying this and
related empirical evaluations of published ITSA are publicly available
in an interrupted time series repository \cite{turnerrepository2025},
which includes numerous examples of outcomes expressed as proportions
or percentages. Boundedness and serial dependence should therefore be
viewed as distinct modeling considerations: addressing one does not
necessarily address the other.

Models designed for proportion outcomes provide a natural starting
point. Fractional regression constrains the conditional mean to the unit
interval and can accommodate observations at zero and one
\cite{papke1996}. For continuous proportions strictly between zero and
one, beta regression provides a distributionally appropriate alternative
\cite{ferrari2004}. When serial dependence is present, beta
autoregressive regression extends beta regression by explicitly modeling
that dependence. A beta-AR($k$) model is the autoregressive special case
of the broader beta-ARMA framework
\cite{benjamin2003,rocha2009,ferreira2015}. Thus, when the dependence
structure can be represented adequately by autoregressive terms,
beta-AR($k$) regression may be used, whereas more complicated temporal
patterns may warrant the broader beta-ARMA model. In either case, the
model addresses the bounded outcome and its temporal dependence within
a common framework.

The purpose of this teaching article is to provide practical guidance
for analyzing proportion outcomes in ITSA by treating boundedness and
serial dependence as separate features that should each be considered
during model selection and evaluation. Two published healthcare applications are reanalyzed to
demonstrate the practical consequences, using data obtained from the
publicly available interrupted time series repository described by Turner
et al.\ \cite{turnerrepository2025}. The first contains observations at
or near both boundaries of the unit interval and illustrates how an
unbounded linear model can produce fitted quantities outside the
permissible range. The second illustrates how serial dependence can
remain unrecognized when autocorrelation is not assessed. Together, the
examples demonstrate a practical approach to selecting, diagnosing, and
evaluating models for proportion-outcome ITSA.

\section{Example 1: When the Outcome Approaches its Boundaries}

Aiken et al.\ evaluated the implementation of a surgical antibiotic
prophylaxis policy at Thika Hospital in Kenya \cite{Aiken2013}. One of
the principal process outcomes was the weekly proportion of patients
undergoing clean or clean-contaminated surgery who received antibiotic
prophylaxis within 60 minutes before surgery. The policy was implemented
in February 2011, with 26 weekly observations before the intervention
and 40 afterward. Preoperative prophylaxis was uncommon during most of the
pre-intervention period, with a sharp increase immediately before the
intervention point, followed by proportions consistently near one
during the post-intervention period. The resulting
series contains observations at or near both boundaries of the unit
interval.

The original study evaluated changes associated with the intervention
using segmented linear regression \cite{Aiken2013}. The mean structure
for a single-group ITSA can be written generally as
\cite{linden2015,linden2017}

\[
g(\mu_t) =
\beta_0 + \beta_1 T_t + \beta_2 X_t
+ \beta_3 (T_t-T_0)X_t ,
\]

where $\mu_t$ is the expected outcome at time $t$, $T_t$ indexes time,
$X_t$ indicates the post-intervention period, and $T_0$ denotes the
intervention time. The coefficients represent the baseline level
($\beta_0$), pre-intervention trend ($\beta_1$), immediate change in
level following the intervention ($\beta_2$), and change in trend
following the intervention ($\beta_3$). The link function $g(\cdot)$
determines how this mean structure is related to the outcome. With
ordinary linear regression, the identity link places no restriction on
the range of the fitted values. A link appropriate for a proportion
outcome can retain the same ITSA mean structure while constraining the
fitted conditional mean to the unit interval.

\begin{figure}[htbp]
\centering
\textbf{(A)}\\[2pt]
\includegraphics[width=0.85\textwidth, keepaspectratio]{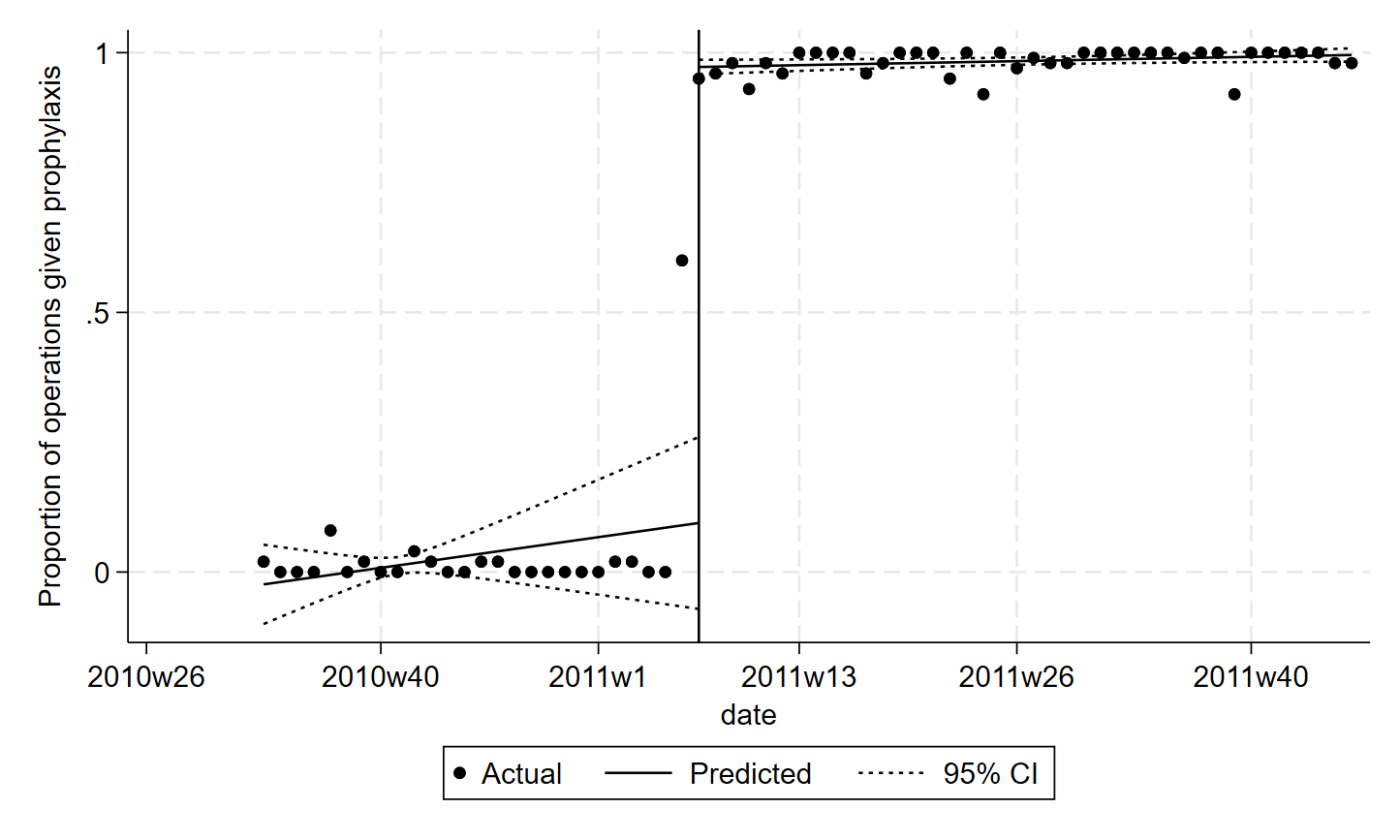}\\[10pt]
\textbf{(B)}\\[2pt]
\includegraphics[width=0.85\textwidth, keepaspectratio]{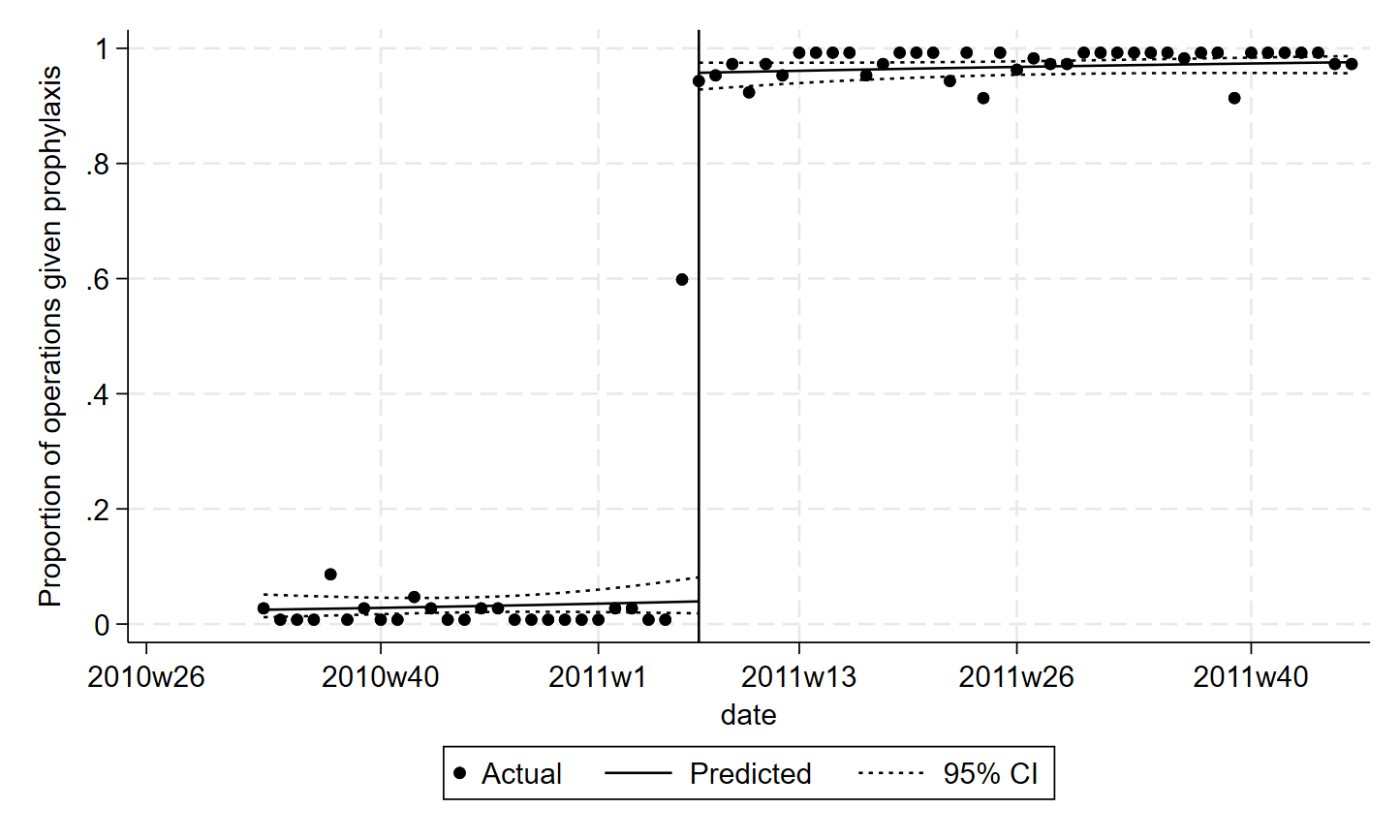}
\caption{Weekly proportion of surgical patients receiving antibiotic prophylaxis at Thika Hospital, Kenya, using data from Aiken et al.~\cite{Aiken2013}. Points show observed proportions; solid lines and dotted lines show fitted values and 95\% confidence intervals from (A) segmented linear regression and (B) fractional logistic regression. The vertical line marks policy implementation (2011w7).}
\label{fig:aiken}
\end{figure}

I first reproduced the original ITSA using linear regression. As shown in
Figure~\ref{fig:aiken}A, the consequences of using an unbounded model
are apparent at both ends of the series. Six fitted values were below
zero. The lower 95\% confidence limit was below zero for all 26
pre-intervention observations, while the upper 95\% confidence limit
exceeded one for the final nine observations. Overall, the 95\%
confidence interval extended outside the permissible $[0,1]$ range at
35 of the 66 time points. These are impossible values for the outcome:
a negative proportion cannot occur, nor can more than 100\% of eligible
patients receive prophylaxis.

I next re-estimated the ITSA using fractional logistic regression
\cite{papke1996}. Fractional regression uses a logit link for the
conditional mean, thereby constraining fitted values to the unit
interval. It also permits observed outcomes equal to zero or one, an
important feature in this application because both boundary values occur
in the series. Standard beta regression, in contrast, assumes a
continuous proportion strictly within $(0,1)$ and therefore cannot be
applied directly when observations equal zero or one
\cite{ferrari2004}.

The fractional logistic model showed no evidence of a
pre-intervention trend ($\beta_1=0.191$, $P=.195$), a large immediate
increase following implementation of the policy
($\beta_2=5.069$, $P=.001$), and no evidence of a change in trend after
the intervention ($\beta_3=-0.152$, $P=.308$). Because these
coefficients are expressed on the logit scale, their magnitudes should
not be interpreted as percentage-point changes. Figure~\ref{fig:aiken}B
therefore presents the fitted trajectory on the proportion scale, where
the fitted values remain within the permissible range.

Serial dependence was also assessed after fitting the fractional
logistic model. The Ljung--Box Portmanteau test showed no evidence of
residual autocorrelation through 12 lags ($Q_{12}=1.45$, $P>.99$)
\cite{ljung1978}. Similarly, the Cumby--Huizinga test
\cite{cumby1992}, implemented using the community-contributed
\texttt{actest} command in Stata \cite{baum2013actest}, showed no
evidence of first-order serial correlation ($\chi^2=0.28$, $P=.600$).
Thus, although the bounded nature of the outcome required attention in
this example, there was no evidence that an autoregressive adjustment
was also needed.

This example does not imply that linear regression will always produce
out-of-range fitted values when the outcome is a proportion. Rather, it
demonstrates that an unbounded model provides no protection against
such values, and that the problem can become consequential when a
series approaches either boundary. It also illustrates why the observed
support of the data matters when choosing among models for proportions.
When legitimate observations of zero or one occur, fractional
regression provides a bounded conditional mean while retaining those
observations unchanged; when a continuous proportion lies strictly
within $(0,1)$, beta regression provides an alternative designed
specifically for such outcomes \cite{ferrari2004}.

Respecting the bounds of the outcome addresses only one of the two
modeling considerations highlighted in this paper. In this example,
assessment of the residuals provided no evidence that serial dependence
also required modeling. The next example illustrates the contrasting
case: an ITSA in which the observed proportions lie within the open
unit interval, but significant autocorrelation is present.

\section{Example 2: When Serial Dependence Is Present}

Barber et al.\ evaluated whether prescribing directives issued by the
Veterans Health Administration (VHA) and the U.S.\ Food and Drug
Administration were associated with changes in opioid prescribing
practices among VHA providers \cite{barber2017}. The study analyzed
monthly proportions of prescriptions meeting specified prescribing
criteria from January 2000 through December 2009. The example considered
here concerns the directive for controlled-release oxycodone, issued in
July 2003. The outcome was the monthly proportion of new
controlled-release oxycodone prescriptions that were preceded by a
prescription for long-acting morphine or methadone. Unlike the first
example, all observed proportions lie strictly within the open unit
interval $(0,1)$, making beta regression applicable.

\begin{figure}[htbp]
\centering

\textbf{(A)}

\includegraphics[width=0.95\textwidth]{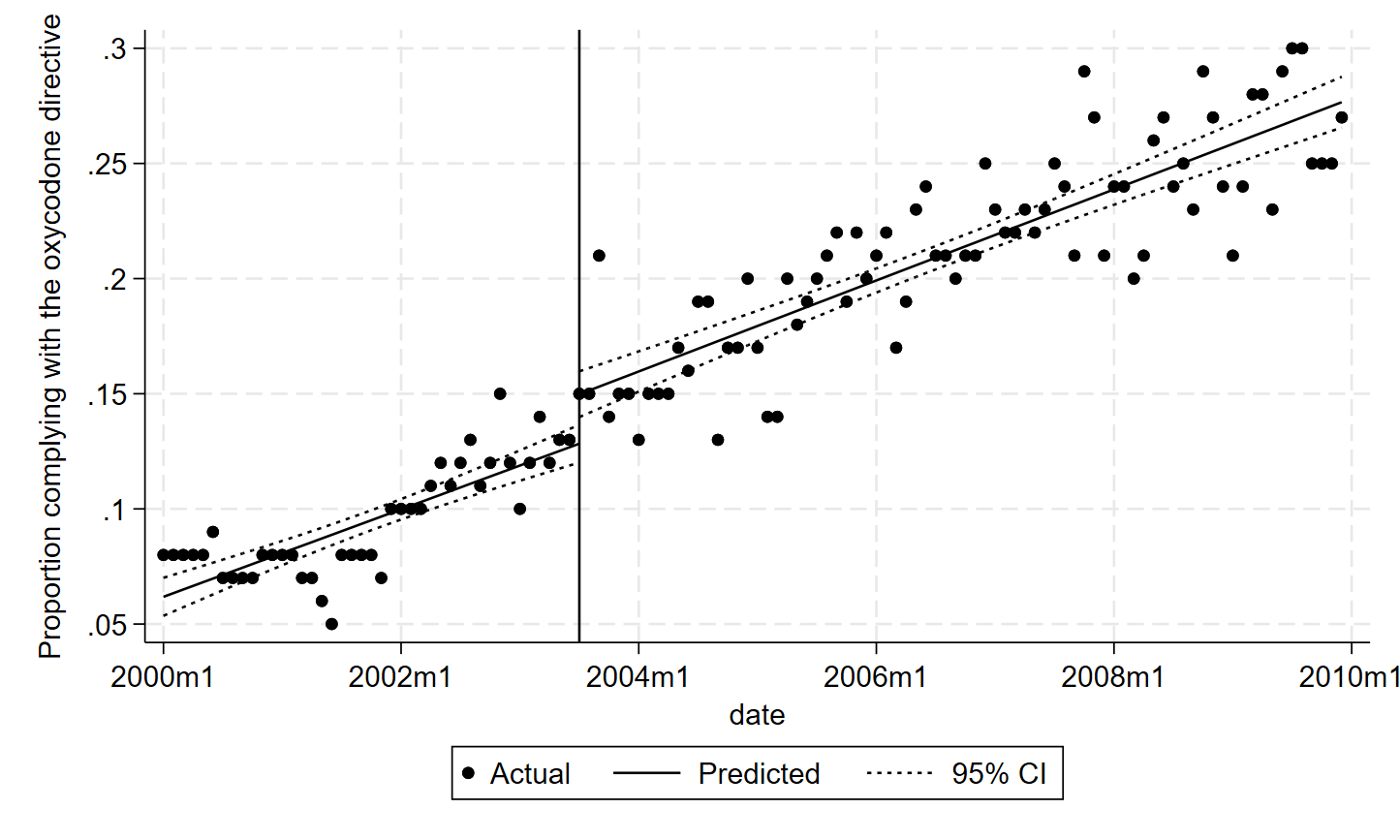}

\vspace{0.5em}

\textbf{(B)}

\includegraphics[width=0.95\textwidth]{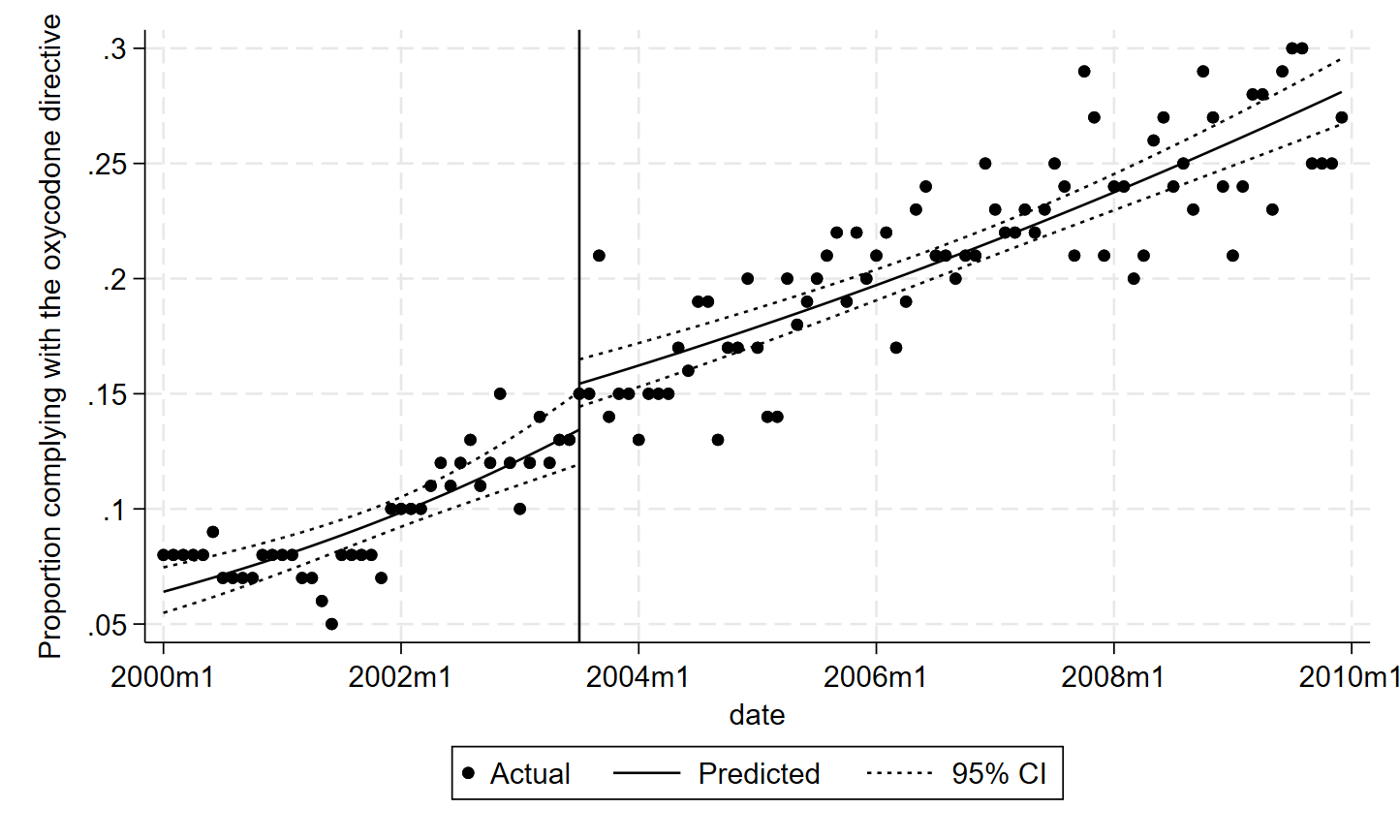}

\caption{Monthly proportion of new controlled-release oxycodone
prescriptions in the Veterans Health Administration preceded by a
prescription for long-acting morphine or methadone, using data from
Barber et al.\ \cite{barber2017}. Points show observed proportions;
solid lines and dotted lines show fitted values and 95\% confidence
intervals from (A) linear regression and (B) beta-AR(1) regression.
The vertical line marks the July 2003 prescribing directive.}

\label{fig:barber}
\end{figure}

The original study analyzed the series using linear ITSA but did not
report an assessment or adjustment for autocorrelation
\cite{barber2017}. I first reproduced the original ITSA using linear
regression (Figure~\ref{fig:barber}A). The fitted trajectory remains
within the permissible range of the outcome and appears reasonable by
visual inspection. Thus, unlike the first example, the principal
modeling problem is not apparent from the fitted trajectory alone.

I next assessed serial dependence before specifying the autoregressive
structure. Examination of the residual autocorrelation function (ACF)
and partial autocorrelation function (PACF) from the unadjusted linear
model indicated positive first-order dependence without a consistent
pattern suggesting a higher-order or seasonal structure. The
Ljung--Box Portmanteau test rejected the null hypothesis of white noise
at lag 1 ($Q_1=4.93$, $P=.026$). Similarly, the Cumby--Huizinga test
identified significant serial correlation at lag 1
($\chi^2=4.81$, $P=.028$), whereas the individual tests for lags 2
through 9 were not significant. Taken together, these diagnostics
supported an AR(1) specification as a parsimonious starting model.

Because the outcome is a continuous proportion within $(0,1)$, the
ITSA was then re-estimated using beta-AR(1) regression, which models
the bounded outcome and first-order serial dependence jointly
\cite{rocha2009,ferreira2015}. Estimation was performed using the
community-contributed \texttt{betark} command in Stata
\cite{lindenbetark}. The beta-AR(1) model estimated a first-order
autoregressive parameter of $\rho_1=0.263$ (95\% CI 0.093--0.432;
$P=.002$), providing further evidence of first-order serial dependence.
The model also estimated a positive pre-intervention trend
($\beta_1=0.0195$, $P<.001$), an immediate increase following the
directive ($\beta_2=0.162$, $P=.040$), and a reduction in the
post-intervention trend relative to the pre-intervention trend
($\beta_3=-0.0096$, $P=.004$). Because these coefficients are
expressed on the logit scale, their magnitudes should not be interpreted
as percentage-point changes. Figure~\ref{fig:barber}B presents the
fitted beta-AR(1) trajectory on the proportion scale.

The two panels of Figure~\ref{fig:barber} illustrate an important
feature of this example. The fitted trajectories from the linear and
beta-AR(1) models are visually similar. Nothing in the linear fit itself
clearly reveals the presence of serial dependence, yet the diagnostic
assessment and beta-AR(1) model provide evidence of first-order
autocorrelation. A fitted trajectory that appears plausible by visual
inspection therefore does not eliminate the need to assess
autocorrelation.

Selection of the autoregressive structure should not end with model
estimation. The residuals should be examined to determine whether the
specified dependence structure has adequately accounted for the serial
correlation. The Ljung--Box Portmanteau test applied to the standardized
conditional residuals from the beta-AR(1) model showed no evidence of
remaining serial dependence ($Q_{12}=15.59$, $P=.211$).
Thus, the postestimation diagnostic supported the
adequacy of the AR(1) specification for these data. Had residual serial
dependence remained, a higher-order beta-AR($k$) specification or the
broader beta-ARMA framework would have warranted consideration
\cite{benjamin2003,rocha2009,ferreira2015}.

This example demonstrates that using a model appropriate for a
proportion outcome does not, by itself, address temporal dependence.
Beta regression respects the bounded support of a continuous proportion
within $(0,1)$, but serial dependence must still be assessed and, when
present, modeled appropriately. The complete process is therefore to
assess the dependence structure, incorporate it into the model, and
then examine the residuals to determine whether the selected structure
was adequate.

\section{Practical Recommendations for Proportion-Outcome ITSA}

Taken together, the two examples illustrate complementary problems in
proportion-outcome ITSA. In the first, the series approached and reached
the boundaries of the unit interval, and an unbounded linear model
produced impossible fitted quantities. In the second, the observed
proportions remained within $(0,1)$ and the fitted linear trajectory
appeared reasonable, but significant serial dependence was present.
Model selection for proportion-outcome ITSA should therefore consider
both the support of the outcome and the temporal dependence among
observations.

Table~\ref{tab:guidance} summarizes a practical framework for model
selection. The first consideration is whether the observed proportion
lies strictly within $(0,1)$ or includes legitimate observations at
zero or one. Continuous proportions strictly within $(0,1)$ can be
modeled using beta regression, whereas fractional regression provides
a convenient alternative when boundary observations are present
\cite{ferrari2004,papke1996}. The second consideration is whether the
series exhibits serial dependence. For an interior proportion with
autoregressive dependence, beta-AR($k$) regression allows the outcome
distribution and dependence structure to be modeled jointly; more
complicated dependence may warrant the broader beta-ARMA framework
\cite{benjamin2003,rocha2009,ferreira2015}.

The final row of Table~\ref{tab:guidance} is intentionally less
prescriptive. The presence of zero or one rules out standard beta
regression, but it does not remove the need to address serial
dependence. Fractional regression can accommodate the boundary values,
and its inference can be made robust to serial dependence using an
appropriate covariance estimator. Alternatively, specialized
time-series models for proportions with boundary values may be
considered when their assumptions and implementation are appropriate.
The important point is that changing the standard errors does not alter
the conditional mean model, just as choosing a bounded conditional mean
does not by itself model serial dependence.

Autocorrelation should be assessed rather than assumed to follow a
particular order. A range of complementary diagnostic approaches is
available, including examination of residual autocorrelation patterns
and formal tests for serial dependence, such as the Ljung--Box
Portmanteau \cite{ljung1978} and Cumby--Huizinga \cite{cumby1992}
tests. No single diagnostic should be regarded as definitive; rather,
the available evidence should be considered together when identifying
a plausible and parsimonious dependence structure. Once a candidate
structure has been incorporated into the model, the residuals should
be reassessed for remaining serial dependence. If residual
autocorrelation persists, a higher-order or otherwise more flexible
dependence structure should be considered. Conversely, additional
complexity is unnecessary when a simpler model adequately accounts for
the observed dependence. This identify--estimate--check sequence is
preferable to routinely assuming AR(1) simply because it is commonly
used in ITSA.

The distinction between modeling serial dependence and adjusting
inference for it is also important. Approaches such as
heteroskedasticity- and autocorrelation-consistent (HAC) standard
errors can adjust estimated uncertainty in the presence of serial
correlation without explicitly modeling the dependence structure.
Simulation evidence comparing fractional regression with HAC standard
errors to beta regression with autoregressive errors suggests that
explicit modeling of the serial dependence can provide better
inferential performance under persistent or higher-order
autocorrelation, although performance depends on the data-generating
conditions \cite{linden2026betaar}. Thus, a covariance correction and
an explicit time-series model should not be regarded as interchangeable
simply because both address consequences of autocorrelation.

The considerations emphasized here do not replace the broader
requirements for specifying a valid ITSA model. Investigators should
also consider features such as seasonality, nonlinear underlying trends,
time-varying confounders, and concurrent events or co-interventions, as
well as whether the intervention effect is expected to be immediate,
delayed, gradual, or otherwise time-varying. These features require separate consideration because neither selecting a model appropriate for the outcome distribution nor accounting for period-to-period serial dependence will, by itself, control for them. More comprehensive guidance on these general aspects of ITSA specification is available elsewhere \cite{linden2015,lopezbernal2017,lopezbernal2018}.

Finally, graphical displays remain useful but should complement rather
than replace diagnostic assessment. In the first example, the
consequences of an unbounded model were immediately apparent because
fitted values and confidence limits crossed the boundaries of the
outcome. In the second, the linear and beta-AR(1) fitted trajectories
were visually similar despite significant first-order serial
dependence. A graph can therefore reveal some forms of model
misspecification while leaving others largely invisible. Assessment of
both the support of the outcome and the residual dependence structure
should be part of routine model evaluation in proportion-outcome ITSA.

\section{Discussion}

The issues considered in this paper are likely to extend beyond the two
examples presented here. Turner et al.\ documented substantial variation
in the statistical methods used in published ITSA studies and found that
autocorrelation was not consistently assessed or reported
\cite{turner2020}. Their accompanying repository contains a large
collection of published interrupted time series datasets, including
series expressed as proportions and percentages
\cite{turnerrepository2025}. Although the present paper was not designed
to systematically evaluate the analytical methods used across that
literature, the availability of these data suggests that the modeling
issues demonstrated here can be explored much more broadly.

The repository also provides a useful resource for extending the
examples in this teaching article. Readers can use the available
proportion and percentage series to examine the range and distribution
of an outcome, compare bounded and unbounded fitted models, investigate
residual serial dependence, and evaluate alternative dependence
structures. Such reanalyses can be used as individual or classroom
learning exercises to explore how modeling choices affect fitted values,
uncertainty, and interpretation using data from actual published studies
rather than simulated examples.

The contribution of this paper is primarily conceptual and
practical rather than the introduction of a new statistical method.
Beta regression, fractional regression, and beta time-series models are
established methods. The central argument is that the distributional
characteristics of the outcome and the temporal dependence of the series
represent separate dimensions of model specification that should be
considered together in proportion-outcome ITSA. This perspective also
helps distinguish methods that explicitly model serial dependence from
approaches that adjust inference for its consequences. For readers
interested in the comparative statistical performance of these
approaches, simulation evidence comparing beta regression with
autoregressive errors and fractional regression with HAC standard errors
is reported elsewhere \cite{linden2026betaar}.

Several limitations should be considered. The two applications were
used as illustrative examples and were not selected through a systematic
review of published proportion-outcome ITSA studies. Accordingly, no
claims are made about the prevalence of inappropriate model
specification in the published literature. The examples also represent
only two configurations of outcome support and temporal dependence and
do not encompass the full range of models that may be appropriate for
bounded time-series data. In particular, proportions containing boundary
values together with complex serial dependence may require methods
beyond those demonstrated here. Finally, the applications use
single-group ITSA. Although the principles concerning bounded outcomes
and temporal dependence are not unique to the single-group design,
additional considerations arise in controlled or multiple-group ITSA \cite{linden2015,linden2017}.

These limitations also point to opportunities for further work. The
public availability of large collections of published ITSA datasets
makes it possible to examine empirically how often proportion outcomes
approach their boundaries, how frequently alternative outcome models
produce materially different fitted values or inference, and how often
serial dependence remains after commonly used specifications. Such work
would complement the illustrative purpose of the present paper and could
help establish how consequential these modeling choices are in applied
ITSA practice.

\section{Conclusions}

For proportion outcomes in ITSA, model specification should reflect both
the support of the outcome and the temporal dependence among observations.
Investigators should select a model appropriate for the observed
proportion, assess the series for serial dependence, and evaluate
residual dependence after estimation. These steps should be incorporated
alongside the broader requirements of good ITSA specification and provide
a practical framework for avoiding preventable model misspecification in
proportion-outcome ITSA.

\section*{Data and code availability}

The data used in both illustrative examples are publicly available from
the interrupted time series repository described by Turner et al.\
\cite{turnerrepository2025}. The \texttt{itsa} command
\cite{linden2015} and the stand-alone \texttt{betark} command
\cite{lindenbetark}, used for the analyses presented here, and the
\texttt{actest} command \cite{baum2013actest}, used for the
Cumby--Huizinga tests, are available from the Statistical Software
Components (SSC) archive.

\bibliographystyle{unsrtnat}
\bibliography{betark_references}

\clearpage

\clearpage

\clearpage
\begin{table}[htbp]
\centering
\caption{Practical considerations for selecting a model for
proportion-outcome ITSA}
\label{tab:guidance}
\begin{tabular}{>{\raggedright\arraybackslash}p{0.25\textwidth} >{\raggedright\arraybackslash}p{0.25\textwidth} >{\raggedright\arraybackslash}p{0.40\textwidth}}
\hline
\textbf{Observed outcome} &
\textbf{Serial dependence} &
\textbf{Modeling consideration} \\
\hline

Continuous proportion strictly within $(0,1)$ &
No evidence of serial dependence &
Beta regression \\
\addlinespace[6pt]

Continuous proportion strictly within $(0,1)$ &
Autoregressive dependence &
Beta-AR($k$) regression \\
\addlinespace[6pt]

Continuous proportion strictly within $(0,1)$ &
More complicated temporal dependence &
Consider the broader beta-ARMA framework \\
\addlinespace[6pt]

Includes legitimate observations at 0 or 1 &
No evidence of serial dependence &
Fractional regression or another model that accommodates boundary
values \\
\addlinespace[6pt]

Includes legitimate observations at 0 or 1 &
Serial dependence present &
Use a boundary-capable proportion model together with an appropriate
method for addressing the temporal dependence; assess model adequacy
after estimation \\
\hline
\end{tabular}
\end{table}

\end{document}